# Evidence of Noncollinear Spin Texture in Magnetic Moiré Superlattices

Hongchao Xie[1,+], Xiangpeng Luo[1,+], Zhipeng Ye[2,+], Gaihua Ye[2], Haiwen Ge[3], Shaohua Yan[4], Yang Fu[4], Shangjie Tian[4], Hechang Lei[4], Kai Sun[1], Rui He[2,*], & Liuyan Zhao[1,*]

[1] Department of Physics, the University of Michigan, 450 Church Street, Ann Arbor, MI, 48109, USA

[2] Department of Electrical and Computer Engineering, Texas Tech University, 910 Boston Avenue, Lubbock, TX, 79409, USA

[3] Department of Mechanical Engineering, Texas Tech University, 2703 7th Street, Lubbock, TX, 79409, USA

[4] Department of Physics and Beijing Key Laboratory of Opto-electronic Functional Materials & Micro-Nano Devices, Renmin University of China, Beijing, 100872, China

[*] Corresponding authors: rui.he@ttu.edu; lyzhao@umich.edu

[+] These authors contributed equally

## ABSTRACT

Moiré magnetism, parallel with moiré electronics that has led to novel correlated and topological electronic states, emerges as a new venue to design and control exotic magnetic phases in twisted magnetic two-dimensional (2D) crystals. Here, we report direct evidence of noncollinear spin texture in 2D twisted double bilayer (tDB) magnet chromium triiodide ($CrI_3$). Using magneto-optical spectroscopy in tDB $CrI_3$, we revealed the presence of a net magnetization, unexpected from the composing antiferromagnetic bilayers with compensated magnetizations, and the emergence of noncollinear spins, originated from the moiré exchange coupling-induced spin frustrations. Exploring the twist angle dependence, we demonstrated that both features are present in tDB $CrI_3$ with twist angles from 0.5º to 5º, but are most prominent in the 1.1º tDB $CrI_3$. Focusing on the temperature dependence of the 1.1º tDB $CrI_3$, we resolved the dramatic suppression in the net magnetization onset temperature and the significant softening of noncollinear spins, as a result of the moiré induced frustration. Our results demonstrate the power of moiré superlattices in introducing novel magnetic phenomena that are absent in natural 2D magnets.

In the presence of magnetic frustrations, where two competing forces for opposite spin alignments cannot be simultaneously satisfied (*1*), noncollinear spin texture can often emerge. Its manifestations include a wealth of exotic magnetic phases in three-dimensional (3D) bulk systems, for example, chiral antiferromagnetism (*2*), spiral magnetism (*3*), helical magnetism (*4*), skyrmions (*5*), magnetic multipoles (*6*) etc. Down to the two-dimensional (2D) limit, noncollinear spin texture has been predicted to develop upon moiré superlattice-induced frustrations, where stacking dependent interlayer exchange coupling with opposite signs coexists within individual moiré supercells and repeats periodically throughout the twisted 2D magnets (*7-9*). But its experimental observation and characterization remain elusive despite very recent efforts in moiré magnetism of twisted chromium triiodide ($CrI_3$) layers (*10-12*).

$CrI_3$ has an out-of-plane easy axis for the ferromagnetic (FM) intralayer exchange coupling (*13*), and thus its long-range magnetic order survives down to the atomic layer limit (*14*). Extensive experiments (*15-17*) and calculations (*18-20*) have shown that monoclinic interlayer stacking leads to antiferromagnetic (AFM) interlayer exchange coupling whereas rhombohedral interlayer atomic registry results in FM interlayer exchange coupling. In natural few-layer $CrI_3$, the structure favors the monoclinic stacking, and consequently the magnetism takes the layered AFM order where spins align along the same out-of-plane direction within the layers but in opposite directions between adjacent layers (*14*). Importantly, even-layer $CrI_3$ has a compensated zero magnetization whereas odd-layer $CrI_3$ ends up with a finite non-zero magnetization. In twisted double odd-layer $CrI_3$, regions with zero and non-zero magnetization coexist at small twist angles (*11, 12*); and in twisted double even-layer $CrI_3$, an intuitively unexpected non-zero magnetization emerges around a critical twist angle of 1.1° (*10*). Yet, the question of spin reorientation, upon moiré modulations, to depart from the collinear arrangement in natural layers remains unexplored in these three pioneering experimental works.

Here, we choose twisted double bilayer (tDB) $CrI_3$ system to show and characterize the noncollinear spins from the moiré engineering. The tDB $CrI_3$ is an artificial four-layer homostructure made by two bilayer (2L) $CrI_3$ stacked vertically with a controlled twist angle α, as shown in the inset of Fig. 1A. At the interface between the two bilayers, a moiré superlattice develops that consists of perfect overlying (AA), rhombohedral (R), and monoclinic (M) stacking within individual moiré supercells (marked as green, blue, and red sites within the black parallelogram, respectively, in Fig. 1A). Based on first principle calculations (*18*), a periodically modulating interlayer exchange coupling of moiré wavelength forms between the 2$^{nd}$ and 3$^{rd}$ layers (*i.e.*, between the two bilayers), with nearly no coupling at the AA sites, ~ 0.1 meV AFM interlayer coupling at the M sites, and ~ – 0.6 meV FM interlayer coupling at R sites, as shown in Fig. 1B. We note that the strength of FM interlayer coupling is significantly stronger than that of AFM.

Using a continuous spin model (*21*), we compute the magnetic energy of the system that includes intralayer FM coupling, uniform interlayer AFM coupling within each bilayer, and moiré modulating interlayer coupling between the two bilayers, and identify the ground state spin configuration by minimizing the magnetic energy. Figures 1C and 1D plot the out-of-plane and in-plane magnetization ($M_z$ and $M_{xy}$) distribution in individual layers, respectively, for the magnetic ground state of tDB CrI$_3$ with a twist angle of α = 1.1°. We find that spatially modulating $M_z$ and $M_{xy}$ form in the 2$^{nd}$ layer of this 1.1° tDB CrI$_3$ that features isolated spin up (↑, $M_z = 3\mu_B$) islands periodically distributed in the spin down (↓, $M_z = -3\mu_B$) background with noncollinear spins (↗↘, $M_{xy} \neq 0$) at the boundaries between these two regions, in contrast to homogeneous spin up, down, and up in the 1$^{st}$, 3$^{rd}$, and 4$^{th}$ layers, respectively. We further correlate the layered magnetism with the lattice stacking geometry in Fig. 1E to show that islands with nonzero magnetization (↑↑↓↑) center at AA sites, boundaries with noncollinear spins coincide with the rings formed by M sites, and background with zero magnetization (↑↓↓↑) are primarily R sites. A linecut of spin orientation in the 2$^{nd}$ layer along the moiré supercell edge (0 – a$_M$) is depicted in Fig. 1F, showing the canted, noncollinear spin texture between the spin up and down regions. The computed results suggest two key features of the moiré magnetism in tDB CrI$_3$: the emergence of an unexpected non-zero magnetization that is consistent with our previous finding (*10*) and the formation of noncollinear spin texture that is the focus of this current study. While the details of spin configurations may vary depending on the microscopic parameters in the model, the presence of a net magnetization and noncollinear spins is robust.

In the following, we use magnetic circular dichroism (MCD) to directly verify the non-zero magnetization and its magnetic field dependence to reveal the noncollinear spin texture (*21*). Figure 2A shows the MCD spectra as a function of an out-of-plane magnetic field (B$_\perp$) swept from + 2T to – 2T and then back to + 2 T, taken on a 1.1° tDB CrI$_3$ (top panel), as well as a 2L (middle panel) and a 4L (bottom panel) CrI$_3$ as reference. Two outstanding distinctions are observed from the comparison between spectra of the 1.1° tDB and natural 2L/4L CrI$_3$, despite similar spin flip transitions at $B_{c1}^{tDB} = \pm 0.67T$ to those at $B_{c1}^{2L} = \pm 0.73T$ (*14*) and $B_{c1}^{4L} = \pm 0.82T$ (*22, 23*). First, in the 1.1° tDB CrI$_3$, the MCD values at $B_\perp = 0T$ are nonzero, and the magnetic field dependence shows a clear hysteresis loop between $B_{c1}^{tDB}$, which is in sharp contrast to the nearly zero MCD values and absence of hysteresis between $B_{c1}^{2L/4L}$ in 2L/4L CrI$_3$. Second, for the 1.1° tDB CrI$_3$, the MCD spectra show gradual increase (i.e., positive slopes) both between and beyond $B_{c1}^{tDB}$, which is clearly distinct from the plateaus of constant MCD values (i.e., zero slopes) outside of the spin flip transition fields for 2L and 4L CrI$_3$. We note that the nonzero MCD at 0T and the hysteresis loop between $B_{c1}^{tDB}$ are strong and direct evidence for a net magnetization in the 1.1° tDB CrI$_3$, as expected from the magnetized islands with a total nonzero $M_z$ in Fig. 1C. We highlight that the gradual increase in MCD outside of $B_{c1}^{tDB}$ suggests the presence of spin flop transitions and thus noncollinear spins in the 1.1° tDB CrI$_3$, consistent with the nonzero $M_{xy}$ at the boundaries between islands and background in Fig. 1D.

We can then model and fit the MCD spectra of the 1.1° tDB CrI$_3$ with two major contributions – one from the collinear spins that feature the sharp spin flip transitions and the other from the noncollinear spins that account for the gradual spin flop transitions (*21*). The fitted results are shown in the top panels of Figs. 2A (solid lines), 2B and 2C for the sum and each of the contributions in the 1.1° tDB CrI$_3$, together with those in the middle and bottom panels for 2L and 4L CrI$_3$, respectively. This fit is independent from but consistent with the computations in Fig. 1 in terms of the following two key aspects. First, the net magnetization falls into the collinear spin contribution in this fit (Fig. 2B, top panel), matching the magnetized islands with collinear ↑↑↓↑ spin arrangements in the computations (Figs. 1C and 1E). Second, the two fitted MCD traces (Fig. 2C, top panel) with decreasing and increasing $B_\perp$ overlap with each other within our fit uncertainties, echoing the computed zero $M_z$, nonzero $M_{xy}$ for the noncollinear spins in Figs. 1C and 1D. As a controlled comparison, the fits for 2L and 4L CrI$_3$ are dominated by the collinear spin contribution with compensated magnetizations below $B_{c1}^{2L/4L}$ (Fig. 2B, middle and bottom panels) and show little noncollinear spin contributions (Fig. 2C, middle and bottom panels).

Based on the computations and the fit, we further provide a physics picture of spin texture evolution upon $B_\perp$ to vividly interpret the MCD spectra. In Fig. 2D, we start with the computed ground state for the 1.1° tDB CrI$_3$ at 0T (step I) and depict the evolution of spin states under increasing $B_\perp$ till 2T. Initially, with $B_\perp$ smaller than $B_{c1}^{tDB}$, the boundaries with noncollinear spins in the 2$^{nd}$ layer gradually move outwards to expand the ↑↑↓↑ islands because the upwards $B_\perp$ direction aligns along the spin up orientation for the island in this 2$^{nd}$ layer (step II). This process accounts for the slow increase of MCD below $B_{c1}^{tDB}$. Right across $B_{c1}^{tDB}$, the collinear spins in the background region flip from ↑↓↓↑ to ↑↑↑↑, which concurrently shifts the modulating spin structure from the 2$^{nd}$ to the 3$^{rd}$ layer (step III). This change, responsible for the sharp jump in MCD at $B_{c1}^{tDB}$, happens because it costs the energy penalty of interlayer exchange coupling with one adjacent layer that $B_{c1}^{tDB}$ can afford. It is similar to the cases of spin flip transitions for 2L CrI$_3$ and for the outer most layer in 4L CrI$_3$ with $B_{c1}^{2L/4L}$ that is close to $B_{c1}^{tDB}$. With further increasing $B_\perp$, the boundaries with noncollinear spins in the 3$^{rd}$ layer move inwards to shrink the ↑↑↓↑ islands where the upwards $B_\perp$ points in the opposite direction to the downwards spins in the islands in this 3$^{rd}$ layer (step IV). This evolution explains the further gradual increase of MCD above $B_{c1}^{tDB}$. Eventually, the islands in the 3$^{rd}$ layer disappear, and spins in all four layers of tDB CrI$_3$ are polarized to the same direction as $B_\perp$ (step V). This final state corresponds to the saturation of MCD, which should happen at $B_\perp$ close to or smaller than $B_{c2}^{4L}$.

Having established the analysis and interpretation of MCD spectra for the 1.1° tDB CrI$_3$, we proceed to explore the twist angle dependence of the moiré magnetism. Figure 3A displays the raw MCD data (top panel) normalized to values at 2T, together with the fits for the sum (top panel) and each (middle and bottom panels) of contributions from collinear and noncollinear spins, for 4L, $\alpha = 0.5°, 1.1°, 2°, 5°, 10°$ tDB, and 2L CrI$_3$ (*21*). A general trend can be seen that the MCD spectra resemble 4L CrI$_3$ at very small twist angles (e.g., 0.5°) in having two spin flip transitions, then distinguish from either 4L or 2L CrI$_3$ at intermediate twist angles (e.g., 1.1°, 2°, and 5°) by showing a significant hysteresis loop and an appreciable noncollinear spin contribution, and eventually converge to 2L at large twist angles (e.g., 10°). To better quantify this trend, we plot the twist angle dependence of four important parameters extracted from the fits, namely, the spin flip transition field $B_{c1}$ that corresponds to the interlayer coupling within each bilayer (Fig. 3B), the width of the spin flip transitions $\Delta B_{c1}$ that describes the spatial inhomogeneity from moiré magnetism (Fig. 3C), the slope at 0T $\Delta MCD/\Delta B_\perp(0T)$ that depicts the susceptibility of the noncollinear spins (Fig. 3D), and the ratio of the noncollinear spin contribution to the MCD $W_{noncollinear}$ that scales with the weight of the noncollinear spins (Fig. 3E). A minimum in $B_{c1}$ at 1.1° in Fig. 3B shows that the moiré interlayer coupling between the two bilayers indeed modifies the uniform interlayer coupling within each bilayer and maximizes this impact at 1.1°. A maximum of $\Delta B_{c1}$ however happens at 0.5° in Fig. 3C, most likely resulted from the extra 4L-like region (↑↓↑↓) within moiré supercells which is marked by the presence of $B_{c2}$ in the 0.5° tDB CrI$_3$ and is proven to be absent in tDB CrI$_3$ of higher twist angles, in addition to the ↑↑↓↑ and ↑↓↓↑ regions. The slope of tDB CrI$_3$ obviously departs from those of 2L and 4L CrI$_3$ and peaks at 1.1°, which is evident of the moiré frustration induced softening for noncollinear spins that lie at the boundaries between collinear spin regions with opposite spin directions. And lastly, the noncollinear spin contribution maximizes at 1.1°, further suggesting the optimal moiré magnetism for tDB CrI$_3$ around this twist angle.

For the optimal twist angle 1.1° tDB CrI$_3$, we further carry out the temperature dependence measurement of MCD spectra from 60 K to 10 K, across the Néel temperature $T_N = 45K$ for natural few-layer CrI$_3$. Figure 4A shows the raw data and fits of MCD spectra at selected temperatures (*21*). At 60K, the MCD spectra shows a linear dependence on $B_\perp$ and confirms the paramagnetic behavior at this temperature. Below 40K, the spin flip transitions at $B_{c1}^{tDB}$ become progressively visible, and furthermore, at 21K, the hysteresis loop with nonzero MCD at 0T appears. We first discuss the fitting parameters for the collinear spin contribution (Figs. 4B–4D). The fitted values for MCD at 0T from the $B_\perp$ dependent MCD spectra kick up at around 22K, which is significantly reduced from the onset of around 45K for the layered AFM in natural 2L/4L CrI$_3$ (*22*) and in each bilayer at R sites in 1.1° tDB CrI$_3$ that is probed by the temperature dependence of layered magnetism-coupled phonon modes in Raman spectra (*10*) (Fig. 4B). This 22K critical temperature results from the development of the moiré superlattice-induced net magnetization in the ↑↑↓↑ islands centering around the AA sites, which is subject to strong moiré frustration and hence results in a dramatic suppression from $T_N$. The temperature dependence of $B_{c1}$ for the 1.1° tDB CrI$_3$ follows the same trend as those for 2L and 4L CrI$_3$ in Fig. 4C, showing an onset around 40K which is consistent with its origin from the AFM interlayer coupling within each bilayers. Yet, the temperature dependence of $\Delta B_{c1}$ for the 1.1° tDB CrI$_3$ clearly departs from those for 2L and 4L CrI$_3$ at temperatures below about 22K approximately when the net magnetization in the ↑↑↓↑ islands builds up. We then move on to the fitting parameters for the noncollinear spin contribution (Figs. 4E–4F). The temperature dependence of the slope peaks (or even diverges) around 40K for both the 1.1° tDB and natural 2L/4L CrI$_3$ (Fig. 4E), resembling that of the AFM susceptibility. Yet, the slope magnitude for the 1.1° tDB CrI$_3$ is much greater than those for 2L and 4L CrI$_3$ right below 40K, evident for the moiré frustration-induced softening of spins. Last but not least, the temperature dependence of $W_{noncollinear}$ below $T_N$ shows a decreasing trend for both 1.1° tDB and 2L/4L CrI$_3$ as the thermal excitation-induced spin canting gets suppressed at lower temperatures. $W_{noncollinear}$ of 1.1° tDB CrI$_3$ is not only much greater than those of 2L and 4L CrI$_3$, but also shows a steeper decrease around 22K when the net magnetization shows up.

We have used magneto-optical measurements of tDB CrI$_3$ to show signatures of an emergent net magnetization and noncollinear spin texture, demonstrate impacts of moiré exchange coupling-induced spin frustrations, and identify the optimal twist angle of around 1.1° for this system. Because our optical studies here are limited to a micrometer resolution and thus only probe the averaged effect across a large number of moiré supercells, future studies by spin-sensitive probes with nanometer spatial resolutions are needed to visualize detailed spin textures within moiré supercells (*24*), and their integration with applications of external magnetic field, if possible, will further enable research on the spin pattern evolution under $B_\perp$. Moreover, dynamic probes, such as time-resolved magneto-optics (*25*), magneto-Raman (*15, 26*), or their near-field versions, or even inelastic x-ray scattering (*27*), are called for to probe moiré magnons in these moiré magnets. Only a combination of static spin textures and dynamic magnons can provide us a complete picture for the moiré magnetism, which can further guide the design and application of interesting magnetic phenomena in moiré magnets.

**Figures and Figure Legends**

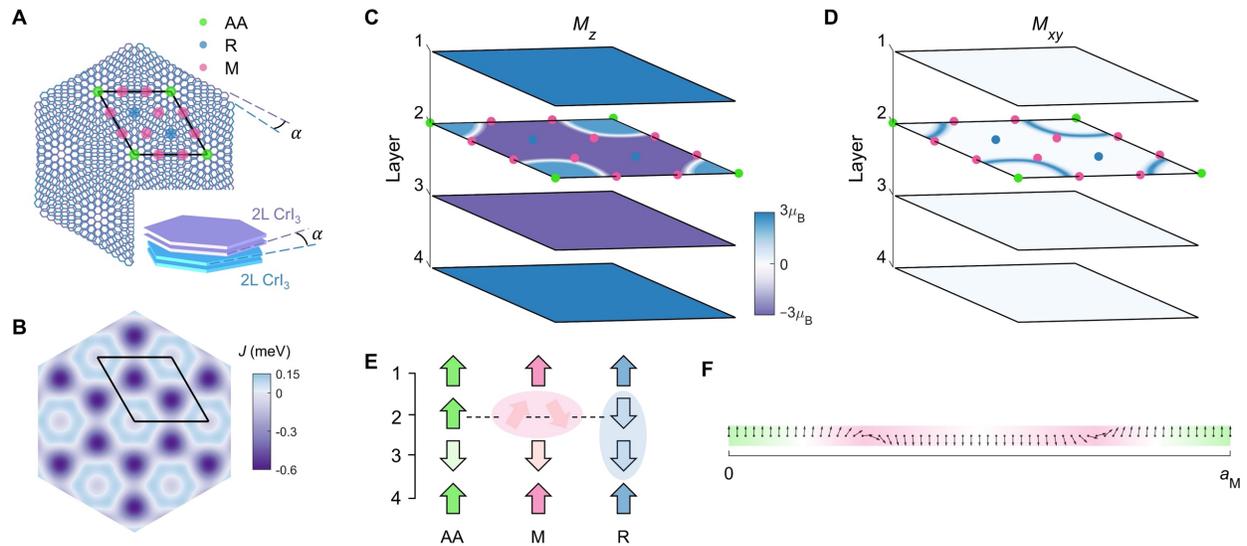

**FIG. 1** Calculations of the magnetic ground state of 1.1° tDB CI$_3$. (**A**) The moiré superlattice formed at the interface between two 2L CrI$_3$ through the interference between the 2$^{nd}$ and 3$^{rd}$ layers. Regions of AA (green), R (blue), and M-type (red) stacking geometries are marked in one moiré supercell (black parallelogram). Inset: schematic of twisting two 2L CrI$_3$ by an angle of α. (**B**) The periodically modulating interlayer exchange coupling J at the interface between two 2L CrI$_3$. (**C** and **D**) The calculated out-of-plane and in-plane magnetization, $M_z$ and $M_{xy}$, distributions in a moiré supercell for all four layers CrI$_3$. (**E**) The sketch of calculated layered magnetism at AA (green), M (red), and R (blue) sites. (**F**) A linecut of spin configuration in the 2$^{nd}$ layer along a moiré supercell edge (0 – $a_M$). Spins belonging to the AA and M sites are shaded with green and red background, respectively.

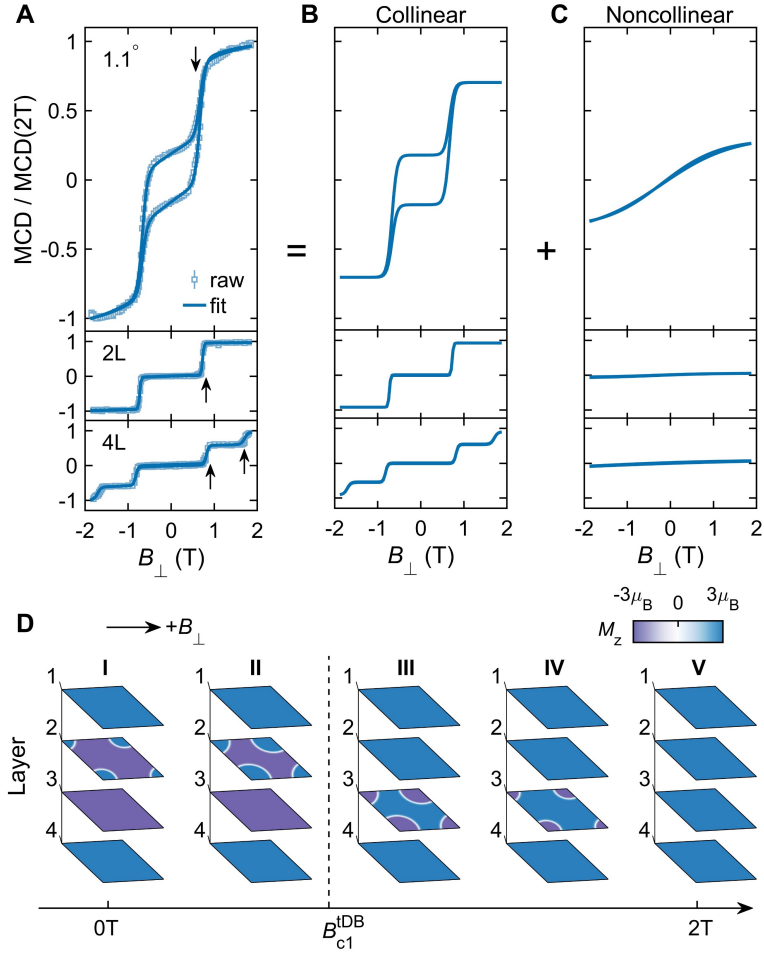

**FIG. 2** Magnetic field dependent MCD spectra, model fits, and spin configurations for 1.1° tDB CrI$_3$. (**A**) MCD spectra and fits normalized to the MCD value at 2T taken at 10 K under an out-of-plane magnetic field $B_\perp$ from + 2T to – 2T and then back to + 2 T, for 1.1° tDB, 2L, and 4L CrI$_3$. The spin flip transitions are marked by the black arrows in all three panels. The error bars correspond to one standard deviation of 15,000 MCD intensities measured over a time interval of 10 s at every magnetic field. The fitting model is detailed in (*21*). (**B** and **C**) Contributions from collinear and noncollinear spins to the MCD extracted from the model fits, in 1.1° tDB, 2L, and 4L CrI$_3$. (**D**) Schematics of $M_z$ distribution evolving upon increasing $B_\perp$ from 0 T across $B_{c1}^{tDB}$ to 2T.

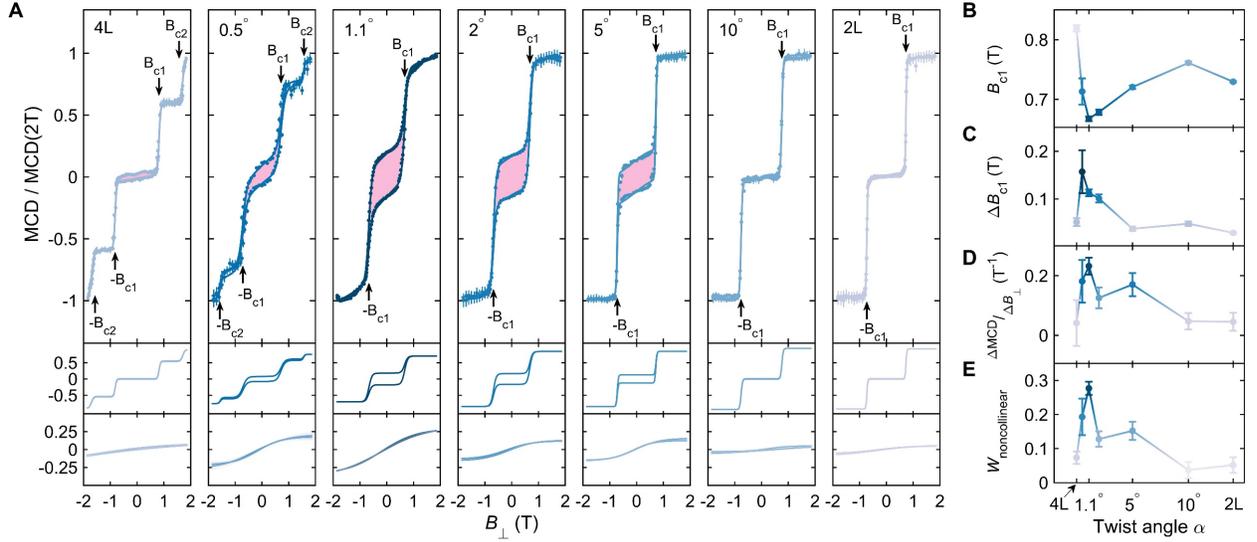

**FIG. 3** Twist angle dependence of MCD data of tDB CrI$_3$. (**A**) Normalized MCD spectra and fits taken at 10 K under $B_\perp$ from +2T to −2T and then back to +2T, for 4L, $\alpha$ = 0.5°, 1.1°, 2°, 5°, 10° tDB, and 2L CrI$_3$ (top panel). The hysteresis loops and spin flip transitions are marked in red shades and with black arrows, respectively. The error bars correspond to one standard deviation of 15,000 MCD intensities measured over a time interval of 10 s at every magnetic field. Contributions from the collinear (middle panel) and the noncollinear (bottom panel) spins are extracted from the model fitting. (**B** – **E**) Twist angle dependence of the spin flip transition field $B_{c1}$, spin flip transition field width $\Delta B_{c1}$, the slope at 0T $\Delta MCD/\Delta B_\perp$, and the weight of the noncollinear spin contribution $W_{\text{noncollinear}}$, extracted from the model fitting. The error bars correspond to two standard error in fitting the MCD spectra.

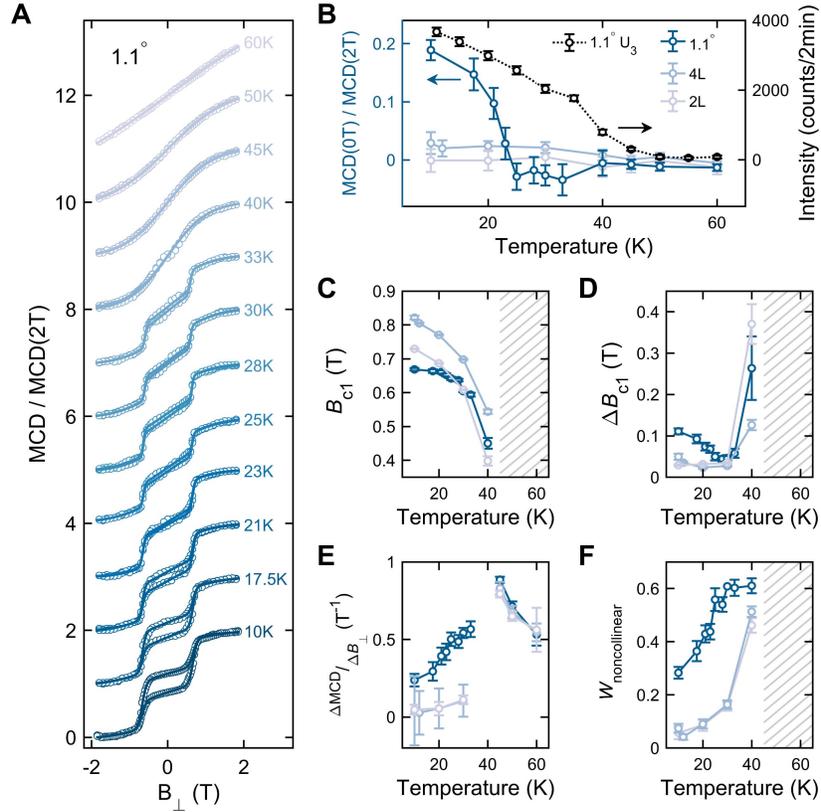

**FIG. 4** Temperature dependence of the MCD data of 1.1° tDB CrI$_3$. (**A**) Normalized MCD spectra and fits at selected temperatures for 1.1° tDB CrI$_3$. Spectra are offset vertically for clarity. (**B** – **F**) Temperature dependence of the normalized MCD value at 0T MCD(0T)/MCD(2T), the spin flip transition field $B_{c1}$, spin flip transition field width $\Delta B_{c1}$, the slope at 0T $\Delta$MCD/$\Delta B_\perp$, and the weight of the noncollinear spin contribution $W_{\text{noncollinear}}$, extracted from the model fitting to MCD data of 1.1° tDB (blue circles), 2L (light purple circles), and 4L (light blue circles) CrI$_3$. Temperature dependence of an antisymmetric Raman mode U$_3$ for the layered AFM order in 1.1° tDB CrI$_3$ is shown in (**B**) (black circles, right y-axis) (*10*). The stripe shaded regions in (**C**, **D**, and **F**) corresponds to temperature ranges where these parameters are absent in the fitting model. The error bars correspond to two standard error in fitting the MCD/Raman spectra.

**Acknowledgements**

**Funding**: L.Z. acknowledges support by NSF CAREER grant no. DMR-174774, AFOSR YIP grant no. FA9550-21-1-0065, and Alfred P. Sloan Foundation. R.H. acknowledges support by NSF grant no. DMR-2104036 and NSF CAREER grant No. 1760668. K.S. acknowledges support by NSF grant no. NSF-EFMA-1741618. H.L. acknowledges support by the National Key R&D Program of China (grant nos. 2018YFE0202600 and 2016YFA0300504), the Beijing Natural Science Foundation (grant no. Z200005), and the Fundamental Research Funds for the Central Universities and Research Funds of Renmin University of China (grant nos. 18XNLG14, 19XNLG17 and 20XNH062).

**Author contributions**: L.Z., H.X., and X.L. conceived the idea and initiated this project. H.X. fabricated the 4L, 2L, and tDB $CrI_3$ samples. H.X., X.L., G.Y., and Z.Y. built the MCD setup and carried out the MCD measurements under the supervision of L.Z. and R. H. S.Y., Y.F., S.T., and H.L. grew the $CrI_3$ bulk single crystals. X.L. and K.S. performed the theoretical computation and analysis. X.L. and L.Z. analyzed the data, and X.L., R.H., and L.Z. wrote the manuscript. All authors participated in the discussion of the results.

**Competing interests**: The authors declare that they have no competing interests.

**Data availability**: All data shown in the main text and supplementary materials are available upon reasonable request from the corresponding authors (R.H. & L.Z.).

**Supplementary Materials**

**This PDF file includes**: Materials and Methods, Supplementary Text, Figs. S1 to S2, and References (*28, 29*).

# Supplementary Materials for

**Evidence of Noncollinear Spin Texture in Magnetic Moiré Superlattices**

Hongchao Xie *et al.*

Correspondence to: Rui He, rui.he@ttu.edu; Liuyan Zhao, lyzhao@umich.edu

**This PDF file includes:**

  Materials and Methods
  Supplementary Text
  Figs. S1 to S2
  References

**Materials and Methods**

Section 1.1    Sample fabrication

The high-quality single crystals of $CrI_3$ were grown by the chemical vapor transport method, as reported in detail in (*28*).

Atomically thin $CrI_3$ films were mechanically exfoliated from bulk crystals on to 300 nm $SiO_2$/Si substrates inside a nitrogen-gas filled glovebox with both $H_2O$ and $O_2$ level below 0.1 ppm. The thickness of $CrI_3$ flakes were initially identified by the optical contrast under an optical microscope inside the glovebox, and then further verified by the Raman spectroscopy at 10 K. Using the "tear and stack" technique, tDB $CrI_3$ samples were fabricated by initially picking up a part of 2L $CrI_3$ and then stacking it on top of the remaining part on the substrate at a targeted twist angle of $\alpha$. Both tDB and 4L/2L $CrI_3$ were encapsulated on both sides with hexagonal boron nitride (hBN) flakes of ~15 nm thick. The final stack of hBN/$CrI_3$/hBN was released onto the $SiO_2$/Si substrates and then rinsed in chloroform solvent to clean the polymer residue, before subsequent magneto-optical measurements.

Section 1.2    MCD measurements

Similar MCD measurements were reported in atomically thin $CrI_3$ flakes in earlier studies (*16*, *17*). For the measurements in this current study, the normal incident light of a wavelength at 632.8 nm was focused onto the hBN/$CrI_3$/hBN sample with a full width at half maximum (FWHM) of 2-3 μm using a ×40 transmissive objective, which is substantially smaller than the sample sizes of both tDB and 4L/2L $CrI_3$. The polarization of the incident light was modulated between left- and right-handed circular polarizations (L/RCP) using the photo-elastic modulator (Hinds Instruments PEM-200) at a modulating frequency $f = 50$ kHz. The reflected signal was collected by a biased photodiode detector. Two separate measurements were carried out to determine the difference between and the average of the LCP and RCP light in the reflection. The difference was measured by demodulating the reflected signal against the PEM frequency using a lock-in amplifier. And the average was measured by inserting a mechanical chopper into the incident beam and then demodulating the reflected light at the mechanical chopper frequency with the lock-in amplifier. The MCD value was defined by the ratio of difference to average signal.

**Supplementary Text**

Section 2.1    Calculations of the magnetic ground state of 1.1° tDB $CrI_3$

We include the intralayer exchange coupling, the spin anisotropy energy, the interlayer AFM exchange coupling within each bilayer, and the modulated interlayer exchange coupling between the two bilayers in the magnetic Hamiltonian

$$H = \sum_{l=1}^{4} H^{(l)} + H_M^{(1,2)} + H_{\text{Moiré}}^{(2,3)} + H_M^{(3,4)}$$

In our calculations, we adopt the continuous spin model. The Moiré supercells are gridded to represent the spin sites, and we use a unit vector $\boldsymbol{n}_i^{(l)}$ to denote the orientation of the $i$-th "spin site" in layer $l$. The intralayer exchange coupling is limited to between the nearest neighbors while the interlayer exchange is only for spins of the same index between two adjacent layers.

For each layer $l$, $H^{(l)}$ summarizes the intralayer FM exchange interaction and the easy-axis anisotropy energy of the spins in this layer

$$H^{(l)} = \frac{1}{\sqrt{3}} J_{\text{intra}} S^2 \sum_{<i,j>} \boldsymbol{n}_i^{(l)} \cdot \boldsymbol{n}_j^{(l)} + 3 J_{\text{intra}} S^2 \sum_{i} (n_{i,x}^{(l)\,2} + n_{i,y}^{(l)\,2} + \alpha n_{i,z}^{(l)\,2})$$

where $J_{\text{intra}} \approx -2.2\text{ meV}/\mu_B^2$ (*18*) is the effective intralayer exchange constant, $S = \frac{3}{2}\mu_B$ is the magnetic moment of Cr$^{3+}$ ions, and $\alpha \approx 1.0445$ (*29*) is the spin anisotropy constant.

$H_M^{(1,2)}$ and $H_M^{(3,4)}$ describes the interlayer exchange interactions within each bilayer–between layers (1,2) and layers (3,4)

$$H_M^{(1,2)} = 2 J_M S^2 \sum_i \boldsymbol{n}_i^{(1)} \cdot \boldsymbol{n}_i^{(2)}$$

$$H_M^{(3,4)} = 2 J_M S^2 \sum_i \boldsymbol{n}_i^{(3)} \cdot \boldsymbol{n}_i^{(4)}$$

where $J_M \approx 0.04\text{ meV}/\mu_B^2$ (*18*) is the effective AFM type interlayer exchange coupling constant between pristine monoclinically-stacked two layers.

$H_{\text{Moiré}}^{(2,3)}$ denotes the periodically modulated magnetic coupling at the twisted interface between layers (2,3)

$$H_{\text{Moire}}^{(2,3)} = 2 J_{\text{Moiré}} S^2 \sum_i \boldsymbol{n}_i^{(2)} \cdot \boldsymbol{n}_i^{(3)}$$

The Moiré modulating interlayer exchange interaction $J_{\text{Moiré}}$ takes the spatial distribution as shown in Figure 1B, adopted from (*18*).

Minimizing the total Hamiltonian with respect to the spin vectors $\boldsymbol{n}_i^{(l)}$ leads to the spin configuration shown in Figs. 1C and 1D. In our calculations, multiple initial conditions were tested to confirm that the noncollinear case in Figs. 1C and 1D is of the lowest energy.

Section 2.2    The fitting model for MCD data

Based on the observation of the MCD spectra, we find that for a typical 1.1° tDB CrI$_3$ measured below $T_N$, MCD spectra consist of two components: 1) the sharp spin flip component with critical fields at $\pm B_{c1}^{\text{tDB}}$ from the collinear spins; 2) the slow-varying component between $-2$ T to $+2$ T from the noncollinear spins.

For the collinear spin contribution, the out-of-plane spins take two possible directions, spin up and down, i.e., $\mu = \pm 2 \cdot \frac{3}{2}\mu_B$. Boltzmann statistics dictates that the mean magnetization under an external field $B$ follows the *tanh* function

$$<\mu> = \frac{3\mu_B e^{\frac{3\mu_B B}{k_B T}} - 3\mu_B e^{-\frac{3\mu_B B}{k_B T}}}{e^{\frac{3\mu_B B}{k_B T}} + e^{-\frac{3\mu_B B}{k_B T}}} \sim \tanh\frac{3\mu_B}{k_B T}B$$

Based on the system equivalency under time-reversal operation for the collinear spins as we learnt from the natural 2L/4L CrI$_3$, the spin-flip transitions adopt two constraints: 1) "flip-up" and "flip-down" transitions should occur at the same magnetic field strength despite opposite magnetic field directions; 2) the increasing-field curve and the decreasing-field curve should map to each other by inversing about the origin at $x = 0$ and $y = 0$ (i.e., $B_\perp = 0$ and MCD = 0 in the MCD spectra).

For the noncollinear spin contribution, phenomenologically we also take a *tanh* function with a wide width to describe the slow-varying feature. Yet, no constraints were added, unlike the collinear spin contributions.

As a result, the following model is used to fit the MCD curve

$$\begin{cases} y_+ = A_{\text{neg}} \tanh\frac{x+B}{C} + A_{\text{pos}} \tanh\frac{x-B}{C} + a_+ \tanh\frac{x-b_+}{c_+} \\ y_- = A_{\text{pos}} \tanh\frac{x+B}{C} + A_{\text{neg}} \tanh\frac{x-B}{C} + a_- \tanh\frac{x-b_-}{c_-} \end{cases}$$

where $x$ is the external magnetic field $B_\perp$, $y_{+(-)}$ is the MCD signal in an increasing(decreasing)-field measurement. For each equation, the first two terms explain the spin-flip transitions of Ising spins and the third term represents the response of the noncollinear spins. $A_{\text{neg(pos)}}$ describes the amplitude of the spin-flip transition, $B$ and $C$ are the transition field $B_{c1}$ and width $\Delta B_{c1}$, respectively. Those parameters are common in fitting $y_+$ and $y_-$ datasets to fulfill the two aforementioned constrains for the collinear spins. For the noncollinear contribution, $a_\pm, b_\pm, c_\pm$ describe the magnitude, center and width of the slow-varying background in either an increasing($+$) or decreasing($-$) measurement, and are independent between the increasing and decreasing field traces.

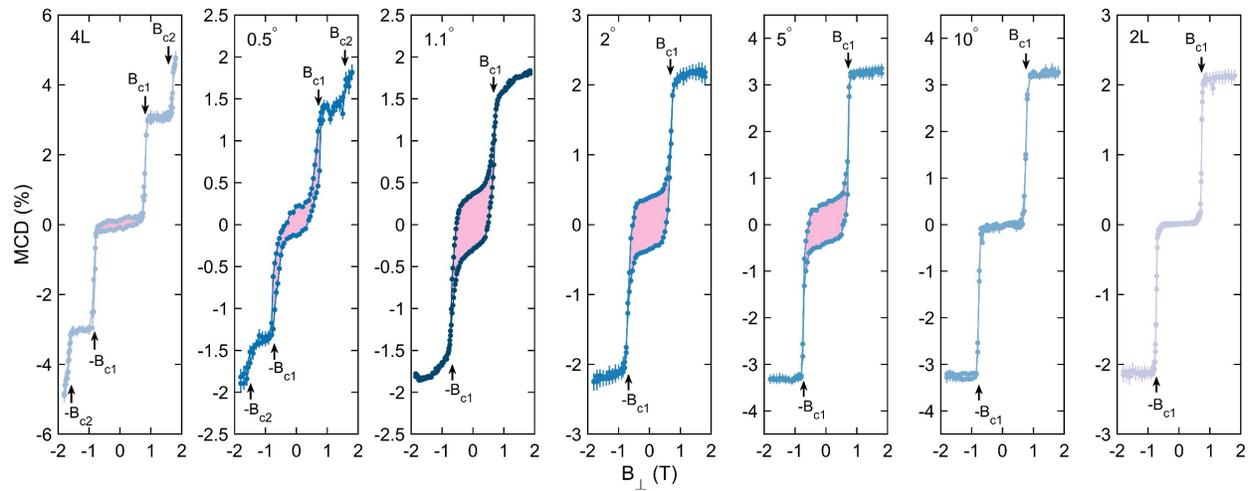

**FIG. S1** Raw MCD data without normalization for 4L, $\alpha$ = 0.5°, 1.1°, 2°, 5°, 10° tDB, and 2L CrI$_3$ samples measured at 10 K. Same data sets as in FIG. 3A.

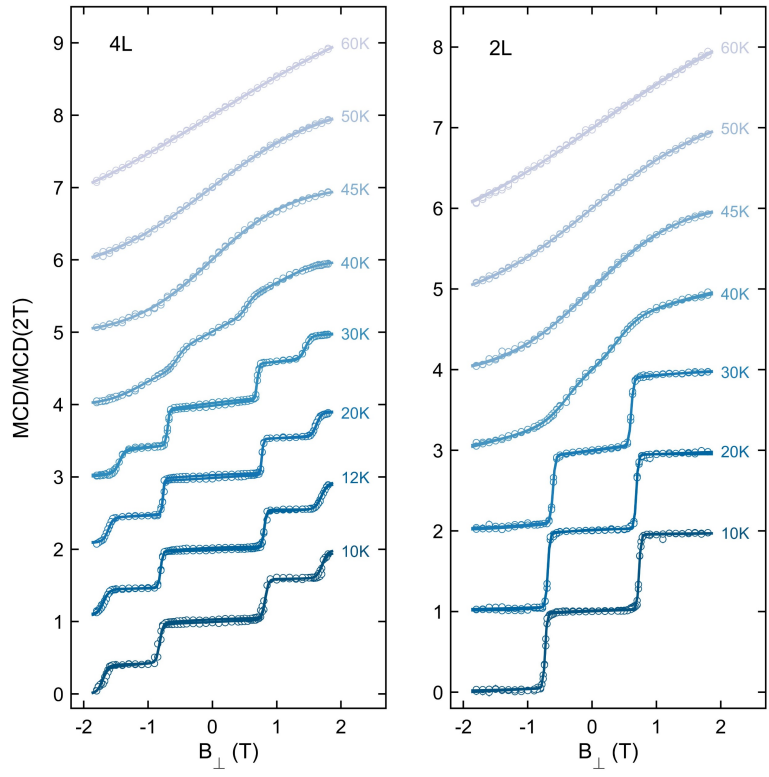

**FIG. S2**  Temperature dependent MCD data and fits for 4L and 2L CrI$_3$ samples. Data sets for the fitting parameters for 4L and 2L CrI$_3$ shown in FIGs. 4B-4F.